\documentclass[reprint, superscriptaddress, amsmath, amssymb, aps, prl]{revtex4-2}
\usepackage{graphicx,dcolumn,bm,xcolor,etoolbox,blindtext,upgreek,float,nicefrac,mathrsfs}
\usepackage[normalem]{ulem}
\usepackage[utf8]{inputenc}
\usepackage[OT1]{fontenc}

\definecolor{prlblue}{RGB}{46, 48, 146}
\definecolor{natcommblue}{RGB}{0, 104, 165}
\usepackage[colorlinks=true,citecolor=prlblue,urlcolor=prlblue,linkcolor=prlblue]{hyperref}
\usepackage{silence}
\newcommand{\dg}{^{\circ}}
\newcommand{\LFO}{Li$_{0.5}$Fe$_{2.5}$O$_4$}
\newcommand{\LAFO}{Li$_{0.5}$Al$_{1.0}$Fe$_{1.5}$O$_4$}
\newcommand{\MGO}{MgGa$_2$O$_4$}
\newcommand{\MAO}{MgAl$_2$O$_4$}
\newcommand{\fref}[2]{\hyperref[#1]{\ref*{#1}#2}}
\newcommand{\sref}[2]{\hyperref[#1]{#2}}

\begin{document}
\title{Divergence between long- and short-wavelength magnon damping in spinel ferrites}

\author{Christopher T. Parzyck}
  \affiliation{Stanford Institute for Materials and Energy Sciences, SLAC National Accelerator Laboratory, 2575 Sand Hill Road, Menlo Park, California 94025, USA}
\author{Octave Duros}
  \affiliation{Stanford Institute for Materials and Energy Sciences, SLAC National Accelerator Laboratory, 2575 Sand Hill Road, Menlo Park, California 94025, USA}
\author{Hari Paudyal}
  \affiliation{Department of Physics and Astronomy, University of Iowa, Iowa City, Iowa 52242, United States}
\author{Noah M. Edmiston}
  \affiliation{Department of Physics, Stanford University, Stanford, California 94305, USA}
  \affiliation{Geballe Laboratory for Advanced Materials, Stanford University, Stanford, California 94305, USA}
\author{Katya Mikhailova}
  \affiliation{Stanford Institute for Materials and Energy Sciences, SLAC National Accelerator Laboratory, 2575 Sand Hill Road, Menlo Park, California 94025, USA}
  \affiliation{Geballe Laboratory for Advanced Materials, Stanford University, Stanford, California 94305, USA}
  \affiliation{Department of Applied Physics, Stanford University, Stanford, California 94305, USA}
\author{Lerato Takana}
  \affiliation{Geballe Laboratory for Advanced Materials, Stanford University, Stanford, California 94305, USA}
  \affiliation{Department of Applied Physics, Stanford University, Stanford, California 94305, USA}
\author{Daisy O’Mahoney}
  \affiliation{Geballe Laboratory for Advanced Materials, Stanford University, Stanford, California 94305, USA}
  \affiliation{Department of Materials Science and Engineering, Stanford University, Stanford, California 94305, USA}
\author{Sauviz P. Alaei}
  \affiliation{Department of Physics, Stanford University, Stanford, California 94305, USA}
  \affiliation{Geballe Laboratory for Advanced Materials, Stanford University, Stanford, California 94305, USA}  
\author{Daniel J. Foster}
  \affiliation{NanoTerasu Promotion Division, Japan Synchrotron Radiation Research Institute (JASRI), 468-1, Aramaki Aza Aoba, Sendai, Miyagi 980-8572, Japan}
\author{Hiroki Suga}
  \affiliation{NanoTerasu Promotion Division, Japan Synchrotron Radiation Research Institute (JASRI), 468-1, Aramaki Aza Aoba, Sendai, Miyagi 980-8572, Japan}
\author{Naoya Kurahashi}
  \affiliation{NanoTerasu Center, National Institutes for Quantum Science and Technology (QST), Miyagi, Japan}
\author{Jun Miyawaki}
  \affiliation{NanoTerasu Center, National Institutes for Quantum Science and Technology (QST), Miyagi, Japan}
\author{Michael E. Flatt\'{e}}
  \affiliation{Department of Physics and Astronomy, University of Iowa, Iowa City, Iowa 52242, United States}
\author{Georgi Dakovski}
  \affiliation{Linac Coherent Light Source, SLAC National Accelerator Laboratory, Menlo Park, CA, 94025, USA}
\author{Yuri Suzuki}
  \affiliation{Stanford Institute for Materials and Energy Sciences, SLAC National Accelerator Laboratory, 2575 Sand Hill Road, Menlo Park, California 94025, USA}
  \affiliation{Geballe Laboratory for Advanced Materials, Stanford University, Stanford, California 94305, USA}
  \affiliation{Department of Applied Physics, Stanford University, Stanford, California 94305, USA}
\author{Durga Paudyal}
  \affiliation{Department of Physics and Astronomy, University of Iowa, Iowa City, Iowa 52242, United States}
\author{Wei-Sheng Lee}
  \affiliation{Stanford Institute for Materials and Energy Sciences, SLAC National Accelerator Laboratory, 2575 Sand Hill Road, Menlo Park, California 94025, USA}

\begin{abstract}
The realization of practical, high-speed magnonic devices requires engineering magnetic materials with low dissipation over wide frequency ranges.  While optical and microwave probes are used to infer the damping of low energy/long wavelength modes, the degree to which these $q\sim0$\ properties translate into higher-energy, finite-momentum modes remains an important open question. Here, we utilize a combination of ferromagnetic resonance (FMR) and resonant inelastic x-ray scattering on spinel ferrites Li$_{0.5}$Al$_x$Fe$_{2.5-x}$O$_4$\ to probe magnons in both the short- and long-wavelength limits.  We observe that aluminum substitution both markedly reduces the magnon bandwidth and drastically shortens the high-\textit{q} magnon lifetimes, in sharp contrast to the ultralow magnon damping inferred from FMR. These findings demonstrate a disparity between how non-magnetic substituents impact magnon damping in the long- and short-wavelength limits, providing a new perspective for assessing candidate materials for magnonic devices.   
\end{abstract} 

\maketitle
Magnonic devices and interconnects, which leverage spin waves (magnons) as energy-efficient information carriers, hold great promise for next-generation, high-speed electronic applications \cite{pirro2021advances,flebusRecentAdvancesMagnonics2023,hanMagnonicsMaterialsPhysics2024}.  Realizing practical magnonic technologies, however, requires development of magnetic materials with minimal magnon dissipation in the GHz and THz ranges. Traditionally, low magnon loss is inferred from the Gilbert damping parameter $\upalpha_g$, measured via ferromagnetic resonance (FMR) \cite{kittelRelaxationProcessFerromagnetism1953,farle1998ferromagnetic} or the spectral linewidths observed in Brillouin light scattering (BLS) \cite{sandercockLightScatteringThermal1973,hillebrandsBrillouinLightScattering1989,demokritov2001brillouin}. 
These optical and microwave techniques primarily probe long-wavelength (momentum $q\sim 0$) excitations near the Brillouin zone center, characterizing magnon dispersion and damping on energy scales of 10's of GHz and momenta up to 10 $\mu$m$^{-1}$\ (0.001 \AA$^{-1}$) \cite{sandwegWiderangeWavevectorSelectivity2010,dunaginBrillouinLightScattering2025}.  Short-wavelength, high-momentum magnons offer potential for much faster propagation speeds essential for high-speed device operation; it is not, however, guaranteed that the ultra-low dissipation observed at (near) zero momentum will extend to finite momenta.
Understanding how material composition, including non-magnetic impurities, for example, impact damping in these two regimes will be critical in the engineering of materials for high-frequency magnonic devices.

Resonant inelastic x-ray scattering (RIXS) has emerged as a powerful tool for probing the low-energy bosonic excitations of materials.  Unlike inelastic neutron scattering (INS), which has been used extensively in the study of bulk low-damping ferrites including yttrium iron garnets, Y$_3$Fe$_5$O$_{12}$\ (YIG) \cite{fergusonScatteringPolarizedNeutrons1967,princepFullMagnonSpectrum2017,nambuNeutronScatteringStudy2021}, the resonantly enhanced cross section and elemental selectivity of RIXS make it an ideal tool to study magnetic excitations not only in bulk ferrites \cite{elnaggarMagneticContrastSpinFlip2019,elnaggarPossibleAbsenceTrimeron2020} but also in thin films \cite{liSingleMultimagnonDynamics2023}, heterostructures \cite{meyersMagnetismIridateHeterostructures2019,shresthaTunableMagnonsAntiferromagnetic2025}, and in out-of-equilibrium systems \cite{guObservingDifferentialSpin2025,Dean2016}.

In this Letter, we simultaneously examine magnetic excitations in the small and large \textit{q} regimes by leveraging both FMR and RIXS to probe the magnetic band structure and damping of insulating thin film spinel ferrites across the Brillouin zone. Through comparison with \textit{ab initio} calculations of the magnon spectrum, we elucidate the impact of non-magnetic dopants on the magnon bandwidth and dispersion.  Finally, we demonstrate that the magnon lifetimes observed at finite momentum may vary drastically, even across materials exhibiting comparable low-loss properties in FMR \cite{zheng2020ultra,omahoney2023aluminum,zhengUltrathinLithiumAluminate2023} and BLS \cite{tongDirectObservationTunable2026}.

\begin{figure*}
  \resizebox{160 mm}{!}{\includegraphics{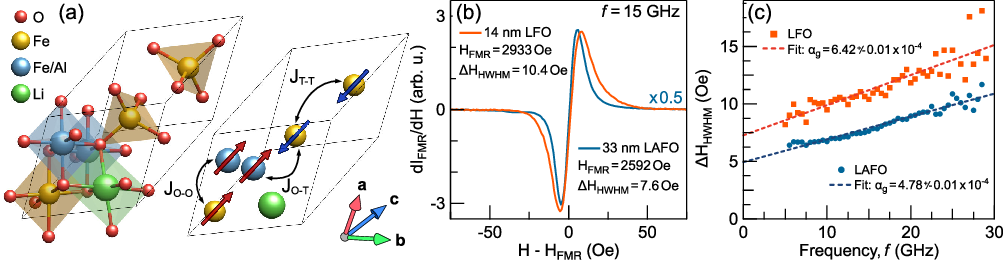}}
  \caption{\label{fig:1}  (a) Crystal and magnetic structure of \LFO\ and \LAFO\ highlighting preferential substitution of Al and Li on the octahedral sites. Moments of the octahedral sites are aligned ferromagnetically by the positive exchange $J_{\textrm{O-O}}$, as are those on the tetrahedral sites by $J_{\textrm{T-T}}$.  Antiferromagnetic exchange, $J_{\textrm{O-T}}$, between the two sites results in overall ferrimagnetic behavior.  (b) Representative 15 GHz FMR traces of LFO and LAFO; the LAFO trace is rescaled by a factor of 0.5. (c) Frequency dependence of the FMR linewidth, $\Delta H_{\textrm{HWHM}}$, and fits from which the Gilbert damping, $\upalpha_g$, is extracted. 
  }
\end{figure*}

We investigate two members of the the family of spinel ferrites, Li$_{0.5}$Al$_x$Fe$_{2.5-x}$O$_4$, which have emerged as promising candidates for magnonic devices owing to their low damping \cite{pachauriStudyStructuralFerromagnetic2015,omahoney2023aluminum,zheng2023ultra,takana2025low}, high Curie temperature \cite{argentinaMicrowaveLithiumFerrites1974}, strain tunable magnetic anisotropy \cite{zhengUltrathinLithiumAluminate2023}, and low synthesis temperatures relative to garnets \cite{omahoney2023aluminum,takana2025low}.  Figure \fref{fig:1}{(a)} shows the inverse spinel crystal structure of Li$_{0.5}$Al$_x$Fe$_{2.5-x}$O$_4$\ with Fe$^{3+}$\ occupying the tetrahedral sites and partially occupying the octahedral sites. In the $x=0$\ member, \LFO\ (LFO), one octahedral site is occupied by Li$^{1+}$ and the other three by Fe$^{3+}$, whereas in the $x=1$ member, \LAFO\ (LAFO), two octahedral sites contain non-magnetic Al$^{3+}$  \cite{whiteMagneticPropertiesLithium1978,argentinaMicrowaveLithiumFerrites1974}. The presence, however, of Al and Li antisite defects are suggested by prior neutron and x-ray studies \cite{naidenNeutronDiffractionLithium1968,omahoney2023aluminum}.  Both LFO and LAFO exhibit room temperature ferrimagnetism \cite{remeikaPropertiesSingleCrystalLithium1964,whiteMagneticPropertiesLithium1978} owing to the strong antiferromagnetic exchange between octahedral and tetrahedral sites, $J_{\textrm{O-T}}$.  

Thin films of LFO and LAFO are grown epitaxially on \MGO~ (MGO) and \MAO~ (MAO) substrates, respectively, by pulsed laser deposition under nominally similar strain states and with comparable static magnetic properties \cite{SIcite}.  The addition of aluminum in LAFO substantially shrinks the lattice parameters, better matching it to commercial MAO substrates \cite{omahoney2023aluminum} which impart a strain of $\upvarepsilon\approx+1\%$. To grow undoped LFO, we select single crystal MGO \cite{galazka2021MGOgrowth} as a suitable substrate, imparting $\upvarepsilon\approx+0.7\%$ of compressive strain.  This choice enables thicker, phase-pure LFO films which remain coherently strained -- avoiding relaxation and phase segregation observed in growths on other substrates: \MAO, SrTiO$_3$, and Al$_2$O$_3$\ \cite{zhangStraintunableMagneticProperties2015,zhengUltrathinLithiumAluminate2023,loukyaStructuralCharacterizationEpitaxial2016a, prietoTailoringLithiumConcentration2024a}.  Consequently, we demonstrate synthesis of high quality LFO films structurally comparable to the LAFO/MAO system and which display strikingly similar dynamic magnetic responses.  As highlighted by the FMR results in Figs. \fref{fig:1}{(b)} and \fref{fig:1}{(c)}, both LFO and LAFO films exhibit sharp resonance behavior in the 10 GHz range and Gilbert dampings, $\upalpha_g$, well below $1\times 10^{-3}$ -- comparable to the gold standard of YIG \cite{dallivykellyInverseSpinHall2013,hauser2016yttrium,jermain2016low,chang2017role}.  Strikingly, the substitution of non-magnetic Al ions does not noticeably degrade the low-frequency/long-wavelength magnon damping.  The fitted $\upalpha_g=4.78\times10^{-4}$\ for the 33 nm LAFO film differs only slightly from that of the 14 nm LFO film, $\upalpha_g=6.42\times10^{-4}$ -- both of which are well below the damping observed in bulk LFO \cite{pachauriStudyStructuralFerromagnetic2015}.

\begin{figure*}
  \resizebox{170 mm}{!}{\includegraphics{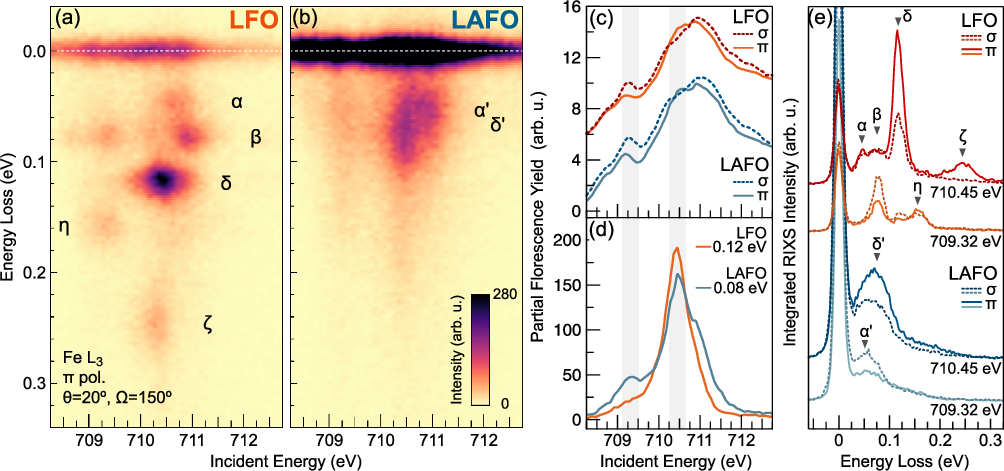}}
  \caption{\label{fig:2} Fe $L_3$-edge RIXS measurements of the LFO and LAFO films characterized in Figs. \fref{fig:1}{(a)} and \fref{fig:1}{(b)}. (a) Incident energy RIXS map of the LFO film taken at $\theta=20\dg$~ grazing incidence with $\pi$ polarized x-rays. (b) RIXS map of the LAFO film taken under identical conditions. (c) Partial fluorescence yield (PFY) x-ray absorption spectrum of the two films obtained by integrating  RIXS maps (a) and (b) over a 13 eV window of energy loss, traces offset for clarity. (d) Resonance profiles of the $\delta$ and $\delta^\prime$ excitations obtained by integrating over a 40 meV energy-loss window about the indicated energies. (e) RIXS spectra obtained by integrating in the incident-energy direction over the 400 meV wide windows indicated by the shaded rectangles in (c) and (d); traces offset for clarity.
  }
\end{figure*}

Having established that the LAFO and LFO films exhibit nearly identical low-damping excitations in the $\sim 10$~GHz (0.04 meV) range, we pivot to an examination of the higher energy ($> 20$~ meV, or 4.8 THz) and momentum magnons by Fe $L_3$-edge RIXS.  Owing to the non-negligible momentum of soft x-rays, RIXS can probe magnetic excitations well away from the zone center. The newly commissioned 2D-RIXS spectrometer at NanoTerasu \cite{yamamotoStatusRIXSBeamline2025,miyawakiAchievingUltrahighResolution2026a} enables both measurement across a wide range of incident energies with high fidelity as well as extremely high resolution.  A RIXS energy map of the LFO film is reported in Fig. \fref{fig:2}{(a)}, taken in grazing incidence with $\pi$\ polarized x-rays.  Modes are visible at five distinct energies, $\alpha:46$, $\beta: 74$, $\delta: 120$, $\eta: 156$, and $\zeta:245$\ meV.  

Modes $\alpha$\ and $\beta$\ are likely attributable to optical phonon modes, falling within the typical energy range associated with phonons in oxide materials ($< 100$~ meV).  Mode $\eta$ exhibits an energy approximately twice that of $\beta$; potentially indicating it is a two-phonon excitation derived from $\beta$. Notably, the sharpest and most intense feature, mode $\delta$, occurs at an energy scale well exceeding that of the typical oxide phonon spectrum.  The relative intensity and energy scale of this feature suggest that it is unlikely to originate from an optical phonon and is instead most consistent with a magnetic excitation.  This conclusion is further supported by its strong momentum-dependent dispersion, which will be discussed later. Mode $\zeta$ appears at an energy approximately twice that of $\delta$, indicating it is potentially a two-magnon excitation similar to multi-magnon features reported in other magnetic insulators \cite{ghiringhelliObservationTwoNondispersive2009,elnaggarMagneticExcitationsSingle2023,liSingleMultimagnonDynamics2023}; we do not presently rule out, however, that it is a high energy optical mode.  In what follows, we will restrict our discussion to the putative single-magnon mode, $\delta$.

Figure \fref{fig:2}{(b)} shows the equivalent RIXS measurement of the Al-substituted LAFO sample.  Here the spectral features are substantially broadened and the peaks corresponding to $\alpha$~ and $\beta$~ are no longer distinguishable.  In the LAFO film, Fig. \fref{fig:2}{(e)}, they instead manifest as a single broad peak, $\alpha'$~ around 44 meV -- consistent with these phonon modes being associated with the octahedral sites where the Fe ions are replaced by Al, resulting in weaker and broader spectral feature. Notably, the strongest excitation $\delta'$, analogous to that of the LFO, exhibits a similar resonance profile, Fig. \fref{fig:2}{(d)}, and polarization dependence, Fig. \fref{fig:2}{(e)}, as the magnetic excitation in LFO. The mode energy is, however, softened substantially to $\delta': 89$~ meV, and is followed by broad remanent spectral weight, which may be associated with the putative two-magnon excitation.  The stark, qualitative differences in the pictures painted by the FMR and RIXS highlights how drastically the magnon response may differ in the large- and small-$q$ regimes.

\begin{figure*}
  \resizebox{170 mm}{!}{\includegraphics{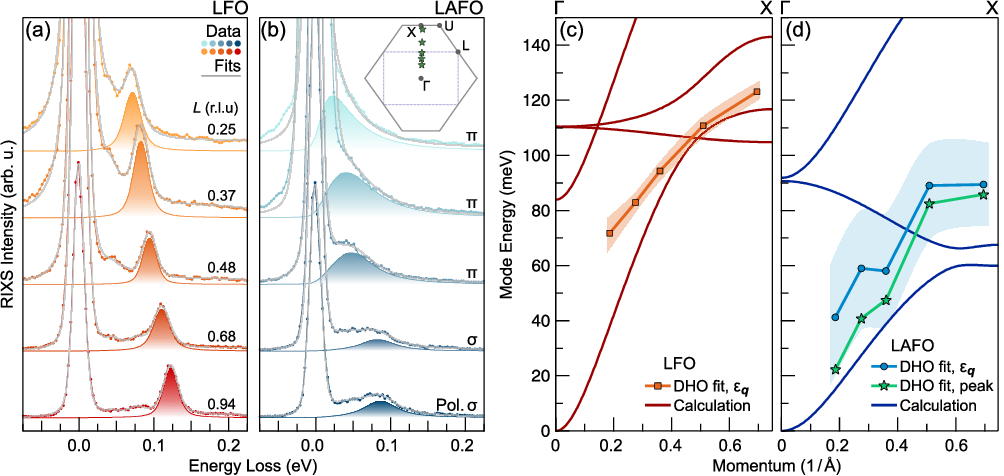}}
  \caption{\label{fig:3} Measured and calculated dispersions of the acoustic magnon in LFO and LAFO. (a) RIXS spectra of the LFO/MGO film along the specular $(001)_{\textrm{cub}}=(\nicefrac{1}{2},\nicefrac{1}{2},0)_{\textrm{Rhom}}$~ direction taken at the magnon resonance energy, 710.45 eV. Quoted $L$~ values are with respect to the measured (pseudo-cubic) c-axis lattice constant of LFO/MGO, 1 r.l.u. = $\nicefrac{2\pi}{c}\approx 0.186$~1/\AA.  (b) Spectra of the LAFO/MAO film taken under the same conditions. (c) Overlay of the fitted magnon energies, $\varepsilon_{\mathbf{q}}$, from (a) with DFT based \textit{ab initio} calculations of the magnon band structure along $\Gamma-\textrm{X}$ direction. (d) The same, for the LAFO/MAO film. The shaded regions in (c) and (d) indicate 95\% confidence intervals in the parameters obtained from the fits.
  }
\end{figure*}

As alluded to above, the magnetic excitations $\delta$ and $\delta'$ display a clear momentum dependence. Figures \fref{fig:3}{(a)} and \fref{fig:3}{(b)} show the RIXS spectra of LFO and LAFO, respectively, measured as a function of momentum transfer, $q$, along the $\Gamma$–$\textrm{X}$ direction at the resonance incident photon energy, $h\nu=710.45$~ eV. The excitation shifts to lower energy as the momentum transfer decreases toward the Brillouin zone center -- consistent with a dispersive mode attributable to the acoustic branch of the magnon spectrum. To extract the magnon band dispersion, at each momentum point the spectra are fit with a damped harmonic oscillator (DHO) model, $\chi''(\mathbf{\hbar\omega,q})$ \cite{pengDispersionDampingIntensity2018} convolved with the instrument's (Gaussian) response function, $\mathrm{FWHM} =2{\sqrt {2\ln 2}}\;\sigma \approx 17$~meV, alongside the elastic response and phonon peaks ($\alpha,\beta,\alpha'$) fixed at the aforementioned energies:
\begin{equation}
    \label{eq:1}
    \chi''(\hbar\omega,\mathbf{q}) = \,\frac{\gamma_{\mathbf{q}}\,\hbar\omega}{\left(\hbar^2\omega^2-\varepsilon_{\mathbf{q}}^2\right)^2+4\gamma_{\mathbf{q}}^2\hbar^2\omega^2}
\end{equation}
where $\hbar\omega$~ is the energy loss, $\mathbf{q}$~ is the momentum, $\varepsilon_{\mathbf{q}}$~ is the mode energy, and $\gamma_{\mathbf{q}}$~ is the magnon damping.  The fitted mode energies are plotted for LFO and LAFO in Figs. \fref{fig:3}{(c)} and \fref{fig:3}{(d)}, respectively.  In the case of LFO, the mode is under-damped, $\gamma_{\mathbf{q}} \ll \varepsilon_{\mathbf{q}}$, and these energies correspond directly to the peak positions of the fit in Fig. \fref{fig:3}{(a)}, shaded.  The mode in LAFO is overdamped, $\gamma_{\mathbf{q}} \gtrsim \varepsilon_{\mathbf{q}}$, however, and as a result the extracted mode energy exceeds that of the fitted peak position, which is also recorded in Fig. \fref{fig:3}{(d)} (stars).  
In this limit, the position of the fitted peaks, Fig. \fref{fig:3}{(b)} (shaded), depend strongly on both $\varepsilon_{\mathbf{q}}$~ and $\gamma_{\mathbf{q}}$, making their accurate determination difficult, particularly at low $q$~ where overlap with the phonons, $\alpha'$, and elastic tail are substantial. For this reason the peak position may serve as a better estimate of the mode energy than the fitted value, $\varepsilon_{\mathbf{q}}$.

Having extracted the experimental magnon dispersions for both LFO and LAFO, we compare them to \textit{ab initio} calculations of the spin-wave dynamics.  In order to capture the competitive interplay of electronic structure driven inter-sub-lattice and intra-sub-lattice exchange pathways, density functional theory (DFT) calculations were performed using the Vienna Ab Initio Simulation Package (VASP)  \cite{kresseEfficiencyAbinitioTotal1996,kresseEfficientIterativeSchemes1996,kresseUltrasoftPseudopotentialsProjector1999} within the Perdew-Burke-Ernzerhof generalized gradient approximation \cite{perdew1996o}, including a mean-field effective Hubbard repulsion $U$. The resulting electronic structure was then downfolded to generate a low-energy localized Hamiltonian \cite{Pizzi2019s}, from which the magnetic super-exchange parameters were extracted \cite{he2021tb2j} and used in a linear spin-wave theory calculation to obtain the magnon bands \cite{Toth2015}.  The results are overlaid on the experimental data in Figs. \fref{fig:3}{(c)} and \fref{fig:3}{(d)} with further details provided in Supplemental Material \cite{SIcite}.  

These calculations show that LFO has a strong ferrimagnetic inter-sub-lattice Fe$^O$-O-Fe$^T$\ super-exchange, $J_{\textrm{O-T}}=-4.26$\ meV, between the tetrahedral and octahedral Fe sites.  It additionally possesses a significant ferromagnetic intra-sub-lattice Fe$^O$-O-Fe$^O$ super-exchange ($J_{\textrm{O-O}}=+1.37$\ meV) with a 92$\dg$\ angle. The interplay between these ferrimagnetic and ferromagnetic interactions governs the low-energy acoustic magnon dispersion, which exhibits both quadratic (near the zone center) and linear (at intermediate momenta) momentum dependencies.  This feature, predicted by the \textit{ab initio} calculations, is observable in the experimentally determined dispersion, which is strongly linear from the lowest measured momentum ($\sim 0.18$\ \AA$^{-1}$) to near the zone boundary.

When non-magnetic Al$^{3+}$\ ions selectively substitute for Fe$^{3+}$~ on the octahedral sites in LAFO, the dilution drastically reduces the density of octahedral Fe pairs.  This severely attenuates the effective ferromagnetic super-exchange $J_{\textrm{O-O}}$ to $+0.09$~ meV, but \textit{not} the localized ferrimagnetic super-exchange, $J_{\textrm{O-T}}=-4.37$~ meV.  This is because the underlying structural geometry and individual Fe$^O$-O-Fe$^T$ super-exchange bond angles $\sim123\dg$~ of LFO and LAFO are similar, preserving the strength of $J_{\textrm{O-T}}$.  The suppression of $J_{\textrm{O-O}}$ in LAFO removes a significant source of magnetic stiffness that otherwise acts cooperatively with $J_{\textrm{O-T}}$ to elevate spin-wave frequencies in LFO. This manifests in the calculated spin-wave dispersion as a distinct softening of both the acoustic and optical magnon branches in LAFO, along with a lowering of the overall spin-wave stiffness constant, $D$.  This behavior is precisely what is observed experimentally with the acoustic branch softening from near 123 meV (LFO) to around 89 meV (LAFO) at the zone edge upon introduction of aluminum.

The DFT-based first-principles calculations well describe several aspects of the observed dispersions, including the overall energy scale of the acoustic mode at large $q$, the linear behavior of the dispersion at intermediate $q$, and the degree of softening induced by the substitution of non-magnetic Al$^{3+}$.  Some quantitative differences between the calculations and RIXS measurements remain, however, such as the overall slope of the acoustic mode as it approaches the Brillouin zone center. Alongside inherent limitations of basic linear spin-wave theory (independent, non-interacting bosons), the quantitative mismatches between the theoretical models and experimental spectra likely stem primarily from variations in the primary exchanges and the omission of higher-order exchange terms; the neglect of antisymmetric Dzyaloshinskii–Moriya interactions may prevent linear spin-wave theory from capturing the directional dependencies and gaps at high-symmetry points. Additionally, magnon-magnon interactions may renormalize the spin-wave energies and the spontaneous decay of high-energy optical branches may shift and broaden the observed states.   

\begin{figure}
  \resizebox{82 mm}{!}{\includegraphics{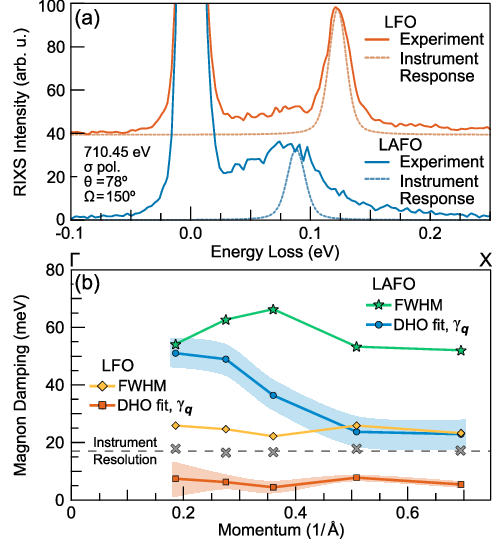}}
  \caption{\label{fig:4}  Comparison of the high-momentum magnon damping between LFO and LAFO. (a) Comparison of the measured magnon spectra (solid) to the instrument response function (dashed) at the same incident energy and momentum transfer.  Magnon damping parameters, $\gamma_{\mathbf{q}}$, obtained by fits to Eqn. \ref{eq:1} along $\Gamma-\textrm{X}$; shaded regions indicate 95\% confidence intervals in the fitted values. The FWHMs of the fitted peaks are also plotted for reference (diamonds, stars), as is the measured instrument resolution (x's).
  }
\end{figure}

While the measured mode energies, $\varepsilon_{\textbf{q}}$, map the high $q$~ dispersion of these Li$^{1+}$ substituted ferrites and reveal the bandwidth tunability induced by the addition of non-magnetic Al$^{3+}$, the spectral linewidths of these peaks also convey information about the mode dampings.  The principal result is the pronounced disparity in the magnon linewidths observed for LAFO relative to LFO. This contrast is illustrated in Fig. \fref{fig:4}{(a)}, where the magnon linewidths of LFO and LAFO in the vicinity of the $\textrm{X}$ are compared to with the RIXS instrument's energy resolution. For LFO, the magnon linewidth is essentially resolution limited, whereas for LAFO it substantially exceeds the instrumental resolution. By fitting to Eqn. \ref{eq:1} the intrinsic damping, $\gamma_{\textbf{q}}$ can be inferred from the experimental data, which are reported in Fig. \fref{fig:4}{(b)}.  In the case of LFO this is straightforward; the fitted values of $\gamma_{\mathbf{q}}$ are essentially identical to the difference between the peak FWHM and the instrument resolution, as is expected in the underdamped case: $\gamma_{\textbf{q}}\lesssim 17\,\textrm{meV}\ll\varepsilon_{\textbf{q}}$. The fitted values of $\gamma_{\textbf{q}}$ for LFO appear constant over the measured range of momenta, around $6.2\pm1$~meV. As discussed above, the case of LAFO is complicated by the large damping $\gamma_{\textbf{q}}\approx\varepsilon_{\textbf{q}}$ and overlap with other spectral features, which disrupts accurate determination of $\gamma_{\textbf{q}}$ especially as $q\rightarrow0$.  In this case the peak FWHM provides an alternative metric for the magnon damping in LAFO, where we find $\gamma\approx\textrm{FWHM}/2=30\pm9$~meV -- markedly above that observed in LFO.  This behavior differs from that of zone-center magnons measured by FMR and BLS, where both LAFO and LFO exhibit low damping and correspondingly narrow linewidths \cite{tongDirectObservationTunable2026}.

Evidently, Al substitution prompts a dramatic increase of the magnon damping at finite momentum, $q>0.1$~\AA$^{-1}$, but has negligible impact on excitations near the Brillouin zone center, $q<0.001$~\AA$^{-1}$. Could disorder associated with Al-substitution explain this observation? Our laboratory based x-ray diffraction measurements indicate both LFO and LAFO exhibit a high degree of crystalline order \cite{SIcite}; however, x-ray magnetic circular dichroism \cite{omahoney2023aluminum} and neutron scattering \cite{naidenNeutronDiffractionLithium1968} measurements of LAFO suggest the addition of aluminum may induce further cation disorder by causing a fraction of the Al$^{3+}$~ and Li$^{1+}$~ ions to migrate to the tetrahedral sites.  Such antisite defects would locally disrupt spin-exchange pathways over a characteristic length scale set by the spatial distribution of the Al and Li ions, plausibly a few to several tens of unit cells. This would be comparable to the magnon wavelengths probed by Fe $L$~ edge RIXS and therefore may induce substantial damping of these finite-$q$ modes.  In contrast, the zone-center magnons measured by BLS and FMR have substantially longer wavelengths, potentially rendering them comparatively insensitive to the local disorder associated with Al substitution.

Another potential source of \textit{q} dependent magnon damping is magnon-phonon coupling, where the crossing of magnon and phonon bands could induce hybridization and increase scattering, reducing the magnon lifetime \cite{andaEffectPhononmagnonInteraction1976}.  The increase in damping associated with this (exchange or anisotropy mediated) interaction is suppressed, however, in the long wavelength limit as $k^{\lambda}$, $\lambda\geq 2$ \cite{woodsMagnonphononEffectsFerromagnetic2001,streibMagnonphononInteractionsMagnetic2019}, making it most relevant to the high \textit{q} modes. The observed magnon-phonon linewidth broadening in INS measurements of manganites \cite{manDirectObservationMagnonphonon2017} and YIG \cite{daiMagnonDampingMagnonphonon2000} measure only in the 5 to 10 meV range -- consistent with the damping of LFO, Fig. \fref{fig:4}{(b)}. However, the magnon linewidth in LAFO appears substantially broader than all three cases, suggesting the presence of additional mechanisms contributing to damping.  It is possible that the bandwidth reduction induced by Al substitution aligns the magnon and phonon density of states more closely in LAFO than in LFO, c.f. Fig. \ref{fig:2}. This overlap -- particularly near the zone boundary, where the magnon density of states peaks -- combined with the substantial magneto-elastic coupling \cite{comstockParallelPumpingMagnetoelastic1964,comstockMagnetoelasticCouplingConstants1965}, could further promote magnon–phonon hybridization for symmetry-allowed modes.  This interpretation is consistent with the pronounced broadening of the phonon features in LAFO observed in Figs. \fref{fig:2}{(b)} and \fref{fig:2}{(e)}. Notably, this coupling would have to be highly mode-selective and predominantly associated with optical phonons involving Fe ionic motion, since the oxygen phonons measured by oxygen K-edge RIXS in this same energy range do not show substantial differences between LFO and LAFO \cite{SIcite}. 

Our case study of LFO and LAFO demonstrates that the low magnon damping observed near the Brillouin zone center ($q \sim 0$) does not necessarily extend to magnons at finite momentum. This divergence highlights the necessity of evaluating magnon damping across both small- and large-\textit{q} regimes when identifying candidate platforms for magnonic applications. Crucially, our findings show that the inclusion of non-magnetic substituents can dramatically alter the magnon dispersion and damping at finite momentum, even though the long-wavelength magnetic properties remain virtually indistinguishable. By leveraging soft x-ray spectroscopy on device relevant thin film materials, we resolve momentum-dependent dynamics beyond the reach of conventional optical and microwave probes and provide complementary insights essential for engineering next-generation, low-loss magnetic materials. \newline

The data that support the findings of this study are available within the paper and supplementary material \cite{SIcite}.  Raw data generated in the course of this study will be made publicly available upon publication through the Stanford Digital Repository. \newline

This work was supported as part of the Center for Energy Efficient Magnonics, an Energy Frontier Research Center funded by the U.S.\@ Department of Energy, Office of Science, Basic Energy Sciences, under Award number DE-AC02-76SF00515. D.P. and H.P. acknowledge the use of the computational facilities on the Frontera supercomputer at the Texas Advanced Computing Center (TACC) via the pathway allocation, DMR23051. The synchrotron radiation experiments were performed at the BL02U of NanoTerasu with the approval of the Japan Synchrotron Radiation Research Institute (JASRI) (Proposal No. 2025B9062). We thank Mr. Rei Tanabe of JASRI for technical support during the measurements. Part of this work was performed at nano@stanford RRID:SCR\_026695.

\bibliography{LAFO}

@article{takana2025low,
  title = {Low Damping (111) Oriented Lithium Aluminum Ferrite Thin Films for Spin Wave Applications},
  author = {Takana, Lerato and Channa, Sanyum and Zheng, Xin Yu and O'Mahoney, Daisy and Alaei, Sauviz and Li, Yuntian and Vailionis, Arturas and Shafer, Padraic and N'Diaye, Alpha T. and Klewe, Christoph and Fisher, Ian and Suzuki, Yuri},
  year = 2025,
  month = jul,
  journal = {Applied Physics Letters},
  volume = {127},
  number = {3},
  pages = {032406},
  doi = {10.1063/5.0278599},
}

@misc{SIcite,
    howpublished = {{See the Supplemental Materials for further information about the synthesis and charictarization of the \LFO~ and \LAFO thin films, details of the first prinicipals magnon calculations, and additional Fe $L$~ and O $K$-edge RIXS measurements}}
}

@article{whiteMagneticPropertiesLithium1978,
    title = {Magnetic properties of lithium ferrite microwave materials},
    volume = {9},
    issn = {0304-8853},
    doi = {10.1016/0304-8853(78)90085-9},
    number = {4},
    journal = {Journal of Magnetism and Magnetic Materials},
    author = {White, G. O. and Patton, C. E.},
    month = dec,
    year = {1978},
    pages = {299--317},
}

@article{remeikaPropertiesSingleCrystalLithium1964,
    title = {Properties of {Single}‐{Crystal} {Lithium} {Ferrite} {Grown} in the {Ordered} {State}},
    volume = {35},
    issn = {0021-8979},
    doi = {10.1063/1.1713215},
    number = {11},
    urldate = {2025-09-24},
    journal = {Journal of Applied Physics},
    author = {Remeika, J. P. and Comstock, R. L.},
    month = nov,
    year = {1964},
    pages = {3320--3321},
}

@article{zhengUltrathinLithiumAluminate2023,
    title = {Ultra-thin lithium aluminate spinel ferrite films with perpendicular magnetic anisotropy and low damping},
    volume = {14},
    issn = {2041-1723},
    url = {https://www.nature.com/articles/s41467-023-40733-9},
    doi = {10.1038/s41467-023-40733-9},
    number = {1},
    urldate = {2024-10-22},
    journal = {Nature Communications},
    author = {Zheng, Xin Yu and Channa, Sanyum and Riddiford, Lauren J. and Wisser, Jacob J. and Mahalingam, Krishnamurthy and Bowers, Cynthia T. and McConney, Michael E. and N’Diaye, Alpha T. and Vailionis, Arturas and Cogulu, Egecan and Ren, Haowen and Galazka, Zbigniew and Kent, Andrew D. and Suzuki, Yuri},
    month = aug,
    year = {2023},
    pages = {4918},
}

@article{omahoney2023aluminum,
  title={Aluminum substitution in low damping epitaxial lithium ferrite films},
  author={O'Mahoney, Daisy and Channa, Sanyum and Zheng, Xin Yu and Vailionis, Arturas and Shafer, Padraic and Klewe, Christoph and Suzuki, Yuri and others},
  journal={Applied Physics Letters},
  volume={123},
  number={17},
  year={2023},
  publisher={AIP Publishing},
  url={https://pubs.aip.org/aip/apl/article/123/17/172405/2918146}
}

@article{Pizzi2019s,
	author = {G. Pizzi and V. Vitale and R. Arita and S. Blügel and F. Freimuth and G. G{\'{e}}ranton and M. Gibertini and D. Gresch and C. Johnson and T. Koretsune and J. Iba{\~{n}}ez-Azpiroz and H. Lee and J.-M. Lihm and D. Marchand and A. Marrazzo and Y. Mokrousov and J. I. Mustafa and Y. Nohara and Y. Nomura and L. Paulatto and S. Ponc{\'{e}} and T. Ponweiser and J. Qiao and F. Thöle and S. S Tsirkin and M. Wierzbowska and N. Marzari and D. Vanderbilt and I. Souza and A. A. Mostofi and J. R. Yates},
	journal = {J. Phys.: Condens. Matter.},
	number = {16},
	pages = {165902},
	title = {Wannier90 as a community code: new features and applications},
	volume = {32},
	year = {2019},
	doi = {10.1088/1361-648X/ab51ff},
	url = {https://iopscience.iop.org/article/10.1088/1361-648X/ab51ff/meta}
}

@article{he2021tb2j,
  title={{TB2J: A python package for computing magnetic interaction parameters}},
  author={He, Xu and Helbig, Nicole and Verstraete, Matthieu J and Bousquet, Eric},
  journal={Comput. Phys. Commun.},
  volume={264},
  pages={107938},
  year={2021},
  publisher={Elsevier},
  url={https://doi.org/10.1016/j.cpc.2021.107938}
}

@article{Toth2015,
  author  = {T{\'o}th, S. and Lake, B.},
  title   = {{Linear spin wave theory for single-Q incommensurate magnetic structures}},
  journal = {J. Phys.: Condens. Matter.},
  volume  = {27},
  number  = {16},
  pages   = {166002},
  year    = {2015},
  doi     = {10.1088/0953-8984/27/16/166002},
  url     = {http://dx.doi.org/10.1088/0953-8984/27/16/166002}
}

@article{pirro2021advances,
  title={Advances in coherent magnonics},
  author={Pirro, Philipp and Vasyuchka, Vitaliy I and Serga, Alexander A and Hillebrands, Burkard},
  journal={Nature Reviews Materials},
  doi={10.1038/s41578-021-00332-w},
  volume={6},
  number={12},
  pages={1114--1135},
  year={2021},
  publisher={Nature Publishing Group UK London}
}

@article{zheng2020ultra,
  title={{Ultra-low magnetic damping in epitaxial Li$_{0.5}$Fe$_{2.5}$O$_4$ thin films}},
  author={Zheng, Xin Yu and Riddiford, Lauren J and Wisser, Jacob J and Emori, Satoru and Suzuki, Yuri},
  journal={Applied Physics Letters},
  doi={10.1063/5.0023077},
  volume={117},
  number={9},
  pages={092407},
  year={2020},
  publisher={AIP Publishing}
}

@article{zheng2023ultra,
  title={Ultra-thin lithium aluminate spinel ferrite films with perpendicular magnetic anisotropy and low damping},
  author={Zheng, Xin Yu and Channa, Sanyum and Riddiford, Lauren J and Wisser, Jacob J and Mahalingam, Krishnamurthy and Bowers, Cynthia T and McConney, Michael E and N’Diaye, Alpha T and Vailionis, Arturas and Cogulu, Egecan and Ren, Haowen and Galazka, Zbigniew and D. Kent, Andrew and Suzuki, Yuri},
  journal={Nature communications},
  doi={10.1038/s41467-023-40733-9},
  volume={14},
  number={1},
  pages={4918},
  year={2023},
  publisher={Nature Publishing Group UK London}
}

@article{farle1998ferromagnetic,
doi = {10.1088/0034-4885/61/7/001},
year = {1998},
month = {jul},
publisher = {},
volume = {61},
number = {7},
pages = {755},
author = {Michael Farle},
title = {Ferromagnetic resonance of ultrathin metallic layers},
journal = {Reports on Progress in Physics},
}

@article{demokritov2001brillouin,
title = {Brillouin light scattering studies of confined spin waves: linear and nonlinear confinement},
journal = {Physics Reports},
volume = {348},
number = {6},
pages = {441-489},
year = {2001},
issn = {0370-1573},
doi = {https://doi.org/10.1016/S0370-1573(00)00116-2},
author = {Demokritov, Sergej O and Hillebrands, Burkard and Slavin, Andrei N}
}

@article{chang2017role,
  title={Role of damping in spin Seebeck effect in yttrium iron garnet thin films},
  author={Chang, Houchen and Praveen Janantha, PA and Ding, Jinjun and Liu, Tao and Cline, Kevin and Gelfand, Joseph N and Li, Wei and Marconi, Mario C and Wu, Mingzhong},
  journal={Science advances},
  doi={10.1126/sciadv.1601614},
  volume={3},
  number={4},
  pages={e1601614},
  year={2017},
  publisher={American Association for the Advancement of Science}
}

@article{hauser2016yttrium,
  title={Yttrium iron garnet thin films with very low damping obtained by recrystallization of amorphous material},
  author={Hauser, Christoph and Richter, Tim and Homonnay, Nico and Eisenschmidt, Christian and Qaid, Mohammad and Deniz, Hakan and Hesse, Dietrich and Sawicki, Maciej and Ebbinghaus, Stefan G and Schmidt, Georg},
  journal={Scientific reports},
  doi={10.1038/srep20827},
  volume={6},
  number={1},
  pages={20827},
  year={2016},
  publisher={Nature Publishing Group UK London}
}

@article{jermain2016low,
    author = {Jermain, C. L. and Paik, H. and Aradhya, S. V. and Buhrman, R. A. and Schlom, D. G. and Ralph, D. C.},
    title = {Low-damping sub-10-nm thin films of lutetium iron garnet grown by molecular-beam epitaxy},
    journal = {Applied Physics Letters},
    volume = {109},
    number = {19},
    pages = {192408},
    year = {2016},
    month = {11},
    doi = {10.1063/1.4967695},
}

@article{perdew1996o,
  title = {{Generalized Gradient Approximation Made Simple}},
  author = {Perdew, J. P. and Burke, K. and Ernzerhof, M.},
  journal = {Phys. Rev. Lett.},
  volume = {77},
  issue = {18},
  pages = {3865--3868},
  numpages = {0},
  year = {1996},
  month = {Oct},
  publisher = {American Physical Society},
  doi = {10.1103/PhysRevLett.77.3865},
  url = {https://link.aps.org/doi/10.1103/PhysRevLett.77.3865}
}

@article{yamamotoStatusRIXSBeamline2025,
    title = {Status of {RIXS} {Beamline} {BL02U} at {NanoTerasu}},
    volume = {3010},
    issn = {1742-6596},
    doi = {10.1088/1742-6596/3010/1/012115},
    number = {1},
    journal = {Journal of Physics: Conference Series},
    publisher = {IOP Publishing},
    author = {Yamamoto, Kohei and Ugalino, Ralph and Fujii, Kentaro and Ohtsubo, Yoshiyuki and Iwasawa, Hideaki and Kitamura, Miho and Imazono, Takashi and Inami, Nobuhito and Nakatani, Takeshi and Inaba, Kento and Agui, Akane and Takeuchi, Tomoyuki and Kimura, Hiroaki and Takahasi, Masamitu and Horiba, Koji and Miyawaki, Jun},
    month = may,
    year = {2025},
    pages = {012115},
}

@article{pengDispersionDampingIntensity2018,
    title = {{Dispersion, damping, and intensity of spin excitations in the monolayer (Bi,Pb)$_2$(Sr,La)$_2$CuO$_{6+\delta}$ cuprate superconductor family}},
    volume = {98},
    doi = {10.1103/PhysRevB.98.144507},
    number = {14},
    urldate = {2026-09-01},
    journal = {Physical Review B},
    publisher = {American Physical Society},
    author = {Peng, Y. Y. and Huang, E. W. and Fumagalli, R. and Minola, M. and Wang, Y. and Sun, X. and Ding, Y. and Kummer, K. and Zhou, X. J. and Brookes, N. B. and Moritz, B. and Braicovich, L. and Devereaux, T. P. and Ghiringhelli, G.},
    month = oct,
    year = {2018},
    pages = {144507},
}

@article{tongDirectObservationTunable2026,
    title = {Direct {Observation} of {Tunable} {Magnons} in {Epitaxial} {Lithium} {Aluminum} {Ferrite} {Thin} {Films}},
    volume = {26},
    issn = {1530-6984},
    url = {https://doi.org/10.1021/acs.nanolett.5c05922},
    doi = {10.1021/acs.nanolett.5c05922},
    number = {9},
    urldate = {2026-08-24},
    journal = {Nano Letters},
    author = {Tong, Junwei and Paudyal, Hari and Liu, Xiangcheng and Takana, Lerato and Mikhailova, Katya and Suzuki, Yuri and Paudyal, Durga and Li, Xiaoqin},
    month = feb,
    year = {2026},
    pages = {3073--3079},
}

@article{sandwegWiderangeWavevectorSelectivity2010,
    title = {Wide-range wavevector selectivity of magnon gases in {Brillouin} light scattering spectroscopy},
    volume = {81},
    issn = {0034-6748},
    url = {https://doi.org/10.1063/1.3454918},
    doi = {10.1063/1.3454918},
    number = {7},
    urldate = {2026-09-07},
    journal = {Review of Scientific Instruments},
    author = {Sandweg, C. W. and Jungfleisch, M. B. and Vasyuchka, V. I. and Serga, A. A. and Clausen, P. and Schultheiss, H. and Hillebrands, B. and Kreisel, A. and Kopietz, P.},
    month = jul,
    year = {2010},
    pages = {073902},
}

@article{pachauriStudyStructuralFerromagnetic2015,
    title = {Study of structural and ferromagnetic resonance properties of spinel lithium ferrite ({LiFe$_5$O$_8$}) single crystals},
    volume = {117},
    issn = {0021-8979},
    url = {https://doi.org/10.1063/1.4922778},
    doi = {10.1063/1.4922778},
    number = {23},
    urldate = {2026-09-04},
    journal = {Journal of Applied Physics},
    author = {Pachauri, Neha and Khodadadi, Behrouz and Althammer, Matthias and Singh, Amit V. and Loukya, B. and Datta, Ranjan and Iliev, Milko and Bezmaternykh, Leonard and Gudim, Irina and Mewes, Tim and Gupta, Arunava},
    month = jun,
    year = {2015},
    pages = {233907},
}

@article{argentinaMicrowaveLithiumFerrites1974,
    title = {Microwave {Lithium} {Ferrites}: {An} {Overview}},
    volume = {22},
    issn = {1557-9670},
    shorttitle = {Microwave {Lithium} {Ferrites}},
    url = {https://ieeexplore.ieee.org/document/1128308},
    doi = {10.1109/TMTT.1974.1128308},
    number = {6},
    urldate = {2026-09-07},
    journal = {IEEE Transactions on Microwave Theory and Techniques},
    author = {Argentina, G.M. and Baba, P.D.},
    month = jun,
    year = {1974},
    pages = {652--658},
}

@article{zhangStraintunableMagneticProperties2015,
    title = {Strain-tunable magnetic properties of epitaxial lithium ferrite thin film on {MgAl$_2$O$_4$} substrates},
    volume = {3},
    issn = {2050-7526},
    doi = {10.1039/c5tc00099h},
    number = {21},
    urldate = {2026-09-01},
    journal = {Journal of Materials Chemistry C},
    author = {Zhang, Ruyi and Liu, Ming and Lu, Lu and Mi, Shao-Bo and Wang, Hong},
    month = jun,
    year = {2015},
    pages = {5598--5602},
}

@article{loukyaStructuralCharacterizationEpitaxial2016a,
    title = {Structural characterization of epitaxial {LiFe$_5$O$_8$} thin films grown by chemical vapor deposition},
    volume = {668},
    issn = {0925-8388},
    doi = {10.1016/j.jallcom.2016.01.217},
    urldate = {2026-09-07},
    journal = {Journal of Alloys and Compounds},
    author = {Loukya, B. and Negi, D. S. and Sahu, R. and Pachauri, N. and Gupta, A. and Datta, R.},
    month = may,
    year = {2016},
    pages = {187--193},
}

@article{prietoTailoringLithiumConcentration2024a,
    title = {Tailoring the {Lithium} {Concentration} in {Thin} {Lithium} {Ferrite} {Films} {Obtained} by {Dual} {Ion} {Beam} {Sputtering}},
    volume = {14},
    copyright = {http://creativecommons.org/licenses/by/3.0/},
    issn = {2079-4991},
    doi = {10.3390/nano14141220},
    number = {14},
    urldate = {2026-09-07},
    journal = {Nanomaterials},
    publisher = {Multidisciplinary Digital Publishing Institute},
    author = {Prieto, Pilar and Hernández-Gómez, Cayetano and Román-Sánchez, Sara and París-Ogáyar, Marina and Gorni, Giulio and Prieto, José Emilio and Serrano, Aida},
    month = jan,
    year = {2024},
    pages = {1220},
}

@article{liSingleMultimagnonDynamics2023,
    title = {Single- and {Multimagnon} {Dynamics} in {Antiferromagnetic} {$\alpha$-Fe$_2$O$_3$}~ {Thin} {Films}},
    volume = {13},
    issn = {2160-3308},
    doi = {10.1103/PhysRevX.13.011012},
    number = {1},
    urldate = {2025-02-13},
    journal = {Physical Review X},
    author = {Li, Jiemin and Gu, Yanhong and Takahashi, Yoshihiro and Higashi, Keisuke and Kim, Taehun and Cheng, Yang and Yang, Fengyuan and Kuneš, Jan and Pelliciari, Jonathan and Hariki, Atsushi and Bisogni, Valentina},
    month = feb,
    year = {2023},
    pages = {011012},
}

@article{elnaggarMagneticExcitationsSingle2023,
    title = {Magnetic excitations beyond the single- and double-magnons},
    volume = {14},
    issn = {2041-1723},
    url = {https://www.nature.com/articles/s41467-023-38341-8},
    doi = {10.1038/s41467-023-38341-8},
    number = {1},
    urldate = {2025-02-17},
    journal = {Nature Communications},
    author = {Elnaggar, Hebatalla and Nag, Abhishek and Haverkort, Maurits W. and Garcia-Fernandez, Mirian and Walters, Andrew and Wang, Ru-Pan and Zhou, Ke-Jin and De Groot, Frank},
    month = may,
    year = {2023},
    pages = {2749},
}

@article{ghiringhelliObservationTwoNondispersive2009,
    title = {Observation of {Two} {Nondispersive} {Magnetic} {Excitations} in {NiO} by {Resonant} {Inelastic} {Soft}-{X}-{Ray} {Scattering}},
    volume = {102},
    doi = {10.1103/PhysRevLett.102.027401},
    number = {2},
    journal = {Physical Review Letters},
    publisher = {American Physical Society},
    author = {Ghiringhelli, G. and Piazzalunga, A. and Dallera, C. and Schmitt, T. and Strocov, V. N. and Schlappa, J. and Patthey, L. and Wang, X. and Berger, H. and Grioni, M.},
    month = jan,
    year = {2009},
    pages = {027401},
}

@article{comstockMagnetoelasticCouplingConstants1965,
    title = {Magnetoelastic coupling constants of the ferrites and garnets},
    volume = {53},
    issn = {1558-2256},
    url = {https://ieeexplore.ieee.org/document/1446193},
    doi = {10.1109/PROC.1965.4263},
    number = {10},
    urldate = {2026-09-08},
    journal = {Proceedings of the IEEE},
    author = {Comstock, R.L.},
    month = oct,
    year = {1965},
    pages = {1508--1517},
}

@article{comstockParallelPumpingMagnetoelastic1964,
    title = {Parallel {Pumping} of {Magnetoelastic} {Waves} in {Lithium} {Ferrite}},
    volume = {136},
    url = {https://link.aps.org/doi/10.1103/PhysRev.136.A442},
    doi = {10.1103/PhysRev.136.A442},
    number = {2A},
    urldate = {2026-09-08},
    journal = {Physical Review},
    publisher = {American Physical Society},
    author = {Comstock, R. L. and Nilsen, W. G.},
    month = oct,
    year = {1964},
    pages = {A442--A445},
}

@article{dallivykellyInverseSpinHall2013,
  title = {Inverse Spin {{Hall}} Effect in Nanometer-Thick Yttrium Iron Garnet/{{Pt}} System},
  author = {{d'Allivy Kelly}, O. and Anane, A. and Bernard, R. and Ben Youssef, J. and Hahn, C. and Molpeceres, A H. and Carr{\'e}t{\'e}ro, C. and Jacquet, E. and Deranlot, C. and Bortolotti, P. and Lebourgeois, R. and Mage, J.-C. and {de Loubens}, G. and Klein, O. and Cros, V. and Fert, A.},
  year = {2013},
  month = aug,
  journal = {Applied Physics Letters},
  volume = {103},
  number = {8},
  pages = {082408},
  doi = {10.1063/1.4819157}
}

@article{guObservingDifferentialSpin2025,
  title = {Observing Differential Spin Currents by Resonant Inelastic {{X-ray}} Scattering},
  author = {Gu, Yanhong and Barker, Joseph and Li, Jiemin and Kikkawa, Takashi and Camino, Fernando and Kisslinger, Kim and Sinsheimer, John and Lienhard, Lukas and Bauer, Jackson J. and Ross, Caroline A. and Basov, Dmitri N. and Saitoh, Eiji and Pelliciari, Jonathan and Bauer, Gerrit E. W. and Bisogni, Valentina},
  year = 2025,
  month = sep,
  journal = {Nature},
  volume = {645},
  number = {8082},
  pages = {900--905},
  issn = {1476-4687},
  doi = {10.1038/s41586-025-09488-9},
}

@article{elnaggarMagneticContrastSpinFlip2019,
    title = {Magnetic {Contrast} at {Spin}-{Flip} {Excitations}: {An} {Advanced} {X}-{Ray} {Spectroscopy} {Tool} to {Study} {Magnetic}-{Ordering}},
    volume = {11},
    shorttitle = {Magnetic {Contrast} at {Spin}-{Flip} {Excitations}},
    doi = {10.1021/acsami.9b10196},
    number = {39},
    urldate = {2025-02-14},
    journal = {ACS Applied Materials \& Interfaces},
    author = {Elnaggar, Hebatalla and Wang, Ru-Pan and Lafuerza, Sara and Paris, Eugenio and Tseng, Yi and McNally, Daniel and Komarek, Alexander and Haverkort, Maurits and Sikora, Marcin and Schmitt, Thorsten and De Groot, Frank M. F.},
    month = oct,
    year = {2019},
    pages = {36213--36220},
}

@article{kresseEfficiencyAbinitioTotal1996,
  title = {Efficiency of Ab-Initio Total Energy Calculations for Metals and Semiconductors Using a Plane-Wave Basis Set},
  author = {Kresse, G. and Furthm{\"u}ller, J.},
  year = 1996,
  month = jul,
  journal = {Computational Materials Science},
  volume = {6},
  number = {1},
  pages = {15--50},
  doi = {10.1016/0927-0256(96)00008-0}
}

@article{kresseEfficientIterativeSchemes1996,
  title = {Efficient Iterative Schemes for Ab Initio Total-Energy Calculations Using a Plane-Wave Basis Set},
  author = {Kresse, G. and Furthm{\"u}ller, J.},
  year = 1996,
  month = oct,
  journal = {Physical Review B},
  volume = {54},
  number = {16},
  pages = {11169--11186},
  publisher = {American Physical Society},
  doi = {10.1103/PhysRevB.54.11169}
}

@article{kresseUltrasoftPseudopotentialsProjector1999,
  title = {From Ultrasoft Pseudopotentials to the Projector Augmented-Wave Method},
  author = {Kresse, G. and Joubert, D.},
  year = 1999,
  month = jan,
  journal = {Physical Review B},
  volume = {59},
  number = {3},
  pages = {1758--1775},
  publisher = {American Physical Society},
  doi = {10.1103/PhysRevB.59.1758}
}

@article{naidenNeutronDiffractionLithium1968,
    title = {Neutron diffraction by lithium ferrite-aluminates},
    volume = {11},
    doi = {10.1007/BF00816068},
    number = {11},
    urldate = {2026-09-09},
    journal = {Soviet Physics Journal},
    author = {Naiden, E. P.},
    month = nov,
    year = {1968},
    pages = {68--71},
}

@article{galazka2021MGOgrowth,
title = {Bulk single crystals of {$\beta$}-{Ga$_2$O$_3$} and {Ga-based} spinels as ultra-wide bandgap transparent semiconducting oxides},
journal = {Progress in Crystal Growth and Characterization of Materials},
volume = {67},
number = {1},
pages = {100511},
year = {2021},
doi = {https://doi.org/10.1016/j.pcrysgrow.2020.100511},
author = {Zbigniew Galazka and Steffen Ganschow and Klaus Irmscher and Detlef Klimm and Martin Albrecht and Robert Schewski and Mike Pietsch and Tobias Schulz and Andrea Dittmar and Albert Kwasniewski and Raimund Grueneberg and Saud Bin Anooz and Andreas Popp and Uta Juda and Isabelle M. Hanke and Thomas Schroeder and Matthias Bickermann},
}

@article{flebusRecentAdvancesMagnonics2023,
  title = {Recent Advances in Magnonics},
  author = {Flebus, B. and Rezende, S. M. and Grundler, D. and Barman, A.},
  year = 2023,
  month = apr,
  journal = {Journal of Applied Physics},
  volume = {133},
  number = {16},
  pages = {160401},
  issn = {0021-8979},
  doi = {10.1063/5.0153424}
}

@article{hanMagnonicsMaterialsPhysics2024,
  title = {Magnonics: {{Materials}}, Physics, and Devices},
  shorttitle = {Magnonics},
  author = {Han, Xiufeng and Wu, Hao and Zhang, Tianyi},
  year = 2024,
  month = jul,
  journal = {Applied Physics Letters},
  volume = {125},
  number = {2},
  pages = {020501},
  issn = {0003-6951},
  doi = {10.1063/5.0216094}
}

@article{sandercockLightScatteringThermal1973,
    title = {Light scattering from thermal acoustic magnons in yttrium iron garnet},
    volume = {13},
    doi = {10.1016/0038-1098(73)90276-7},
    number = {10},
    journal = {Solid State Communications},
    author = {Sandercock, J. R. and Wettling, W.},
    month = nov,
    year = {1973},
    pages = {1729--1732},
}

@article{hillebrandsBrillouinLightScattering1989,
    title = {Brillouin light scattering from spin waves in magnetic layers and multilayers},
    volume = {49},
    doi = {10.1007/BF00616984},
    number = {6},
    urldate = {2026-09-14},
    journal = {Applied Physics A},
    author = {Hillebrands, B. and Baumgart, P. and Güntherodt, G.},
    month = dec,
    year = {1989},
    pages = {589--598},
}

@article{dunaginBrillouinLightScattering2025,
    title = {Brillouin light scattering spectroscopy of magnon–phonon thermal spectra of an in-plane magnetized {YIG} film in two-dimensional wavevector space},
    volume = {137},
    issn = {0021-8979},
    doi = {10.1063/5.0251149},
    number = {8},
    urldate = {2026-09-14},
    journal = {Journal of Applied Physics},
    author = {Dunagin, Ryan E. and Serga, Alexander A. and Bozhko, Dmytro A.},
    month = feb,
    year = {2025},
    pages = {083901},
}

@article{kittelRelaxationProcessFerromagnetism1953,
    title = {Relaxation {Process} in {Ferromagnetism}},
    volume = {25},
    doi = {10.1103/RevModPhys.25.233},
    number = {1},
    urldate = {2026-09-14},
    journal = {Reviews of Modern Physics},
    publisher = {American Physical Society},
    author = {Kittel, C. and Abrahams, Elihu},
    month = jan,
    year = {1953},
    pages = {233--238},
}

@article{fergusonScatteringPolarizedNeutrons1967,
    title = {Scattering of {Polarized} {Neutrons} by {Spin} {Waves} in {Magnetite} and {Yttrium} {Iron} {Garnet}},
    volume = {156},
    doi = {10.1103/PhysRev.156.632},
    number = {2},
    urldate = {2026-09-14},
    journal = {Physical Review},
    publisher = {American Physical Society},
    author = {Ferguson, G. A. and Sáenz, A. W.},
    month = apr,
    year = {1967},
    pages = {632--636},
}

@article{princepFullMagnonSpectrum2017,
    title = {The full magnon spectrum of yttrium iron garnet},
    volume = {2},
    copyright = {2017 The Author(s)},
    doi = {10.1038/s41535-017-0067-y},
    number = {1},
    urldate = {2026-01-16},
    journal = {npj Quantum Materials},
    publisher = {Nature Publishing Group},
    author = {Princep, Andrew J. and Ewings, Russell A. and Ward, Simon and Tóth, Sandor and Dubs, Carsten and Prabhakaran, Dharmalingam and Boothroyd, Andrew T.},
    month = nov,
    year = {2017},
    pages = {63},
}

@article{nambuNeutronScatteringStudy2021,
    title = {Neutron {Scattering} {Study} on {Yttrium} {Iron} {Garnet} for {Spintronics}},
    volume = {90},
    issn = {0031-9015, 1347-4073},
    doi = {10.7566/JPSJ.90.081002},
    number = {8},
    urldate = {2026-09-14},
    journal = {Journal of the Physical Society of Japan},
    author = {Nambu, Yusuke and Shamoto, Shin-ichi},
    month = aug,
    year = {2021},
    pages = {081002},
}

@article{elnaggarPossibleAbsenceTrimeron2020,
    title = {Possible absence of trimeron correlations above the {Verwey} temperature in {Fe$_3$O$_4$}},
    volume = {101},
    doi = {10.1103/PhysRevB.101.085107},
    number = {8},
    journal = {Physical Review B},
    publisher = {American Physical Society},
    author = {Elnaggar, H. and Wang, R. and Lafuerza, S. and Paris, E. and Komarek, A. C. and Guo, H. and Tseng, Y. and McNally, D. and Frati, F. and Haverkort, M. W. and Sikora, M. and Schmitt, T. and de Groot, F. M. F.},
    month = feb,
    year = {2020},
    pages = {085107},
}

@article{meyersMagnetismIridateHeterostructures2019,
    title = {Magnetism in iridate heterostructures leveraged by structural distortions},
    volume = {9},
    copyright = {2019 The Author(s)},
    issn = {2045-2322},
    doi = {10.1038/s41598-019-39422-9},
    number = {1},
    journal = {Scientific Reports},
    publisher = {Nature Publishing Group},
    author = {Meyers, D. and Cao, Yue and Fabbris, G. and Robinson, Neil J. and Hao, Lin and Frederick, C. and Traynor, N. and Yang, J. and Lin, Jiaqi and Upton, M. H. and Casa, D. and Kim, Jong-Woo and Gog, T. and Karapetrova, E. and Choi, Yongseong and Haskel, D. and Ryan, P. J. and Horak, Lukas and Liu, X. and Liu, Jian and Dean, M. P. M.},
    month = mar,
    year = {2019},
    pages = {4263},
}

@article{shresthaTunableMagnonsAntiferromagnetic2025,
    title = {Tunable magnons of an antiferromagnetic {Mott} insulator via interfacial metal-insulator transitions},
    volume = {16},
    copyright = {2025 The Author(s)},
    issn = {2041-1723},
    doi = {10.1038/s41467-025-58922-z},
    number = {1},
    urldate = {2026-09-14},
    journal = {Nature Communications},
    publisher = {Nature Publishing Group},
    author = {Shrestha, Sujan and Souri, Maryam and Dietl, Christopher J. and Pärschke, Ekaterina M. and Krautloher, Maximilian and Calderon Ortiz, Gabriel A. and Minola, Matteo and Shi, Xiatong and Boris, Alexander V. and Hwang, Jinwoo and Khaliullin, Giniyat and Cao, Gang and Keimer, Bernhard and Kim, Jong-Woo and Kim, Jungho and Seo, Ambrose},
    month = apr,
    year = {2025},
    pages = {3592},
}

@article{Dean2016,
    title = {Ultrafast energy- and momentum-resolved dynamics of magnetic correlations in the photo-doped {Mott} insulator {Sr$_2$IrO$_4$}},
    volume = {15},
    issn = {1476-1122},
    doi = {10.1038/nmat4641},
    number = {6},
    journal = {Nature Materials},
    author = {Dean, M P M and Cao, Y and Liu, X. and Wall, S and Zhu, D and Mankowsky, R and Thampy, V and Chen, X M and Vale, J G and Casa, D and Kim, Jungho and Said, A H and Juhas, P and Alonso-Mori, R and Glownia, J M and Robert, A and Robinson, J and Sikorski, M and Song, S and Kozina, M and Lemke, H and Patthey, L and Owada, S and Katayama, T and Yabashi, M and Tanaka, Yoshikazu and Togashi, T and Liu, J and Rayan Serrao, C. and Kim, B J and Huber, L and Chang, C.-L and McMorrow, D. F. and Först, M and Hill, J P},
    month = jun,
    year = {2016},
    pages = {601--605},
}

@article{daiMagnonDampingMagnonphonon2000,
    title = {Magnon damping by magnon-phonon coupling in manganese perovskites},
    volume = {61},
    doi = {10.1103/PhysRevB.61.9553},
    number = {14},
    urldate = {2026-09-14},
    journal = {Physical Review B},
    publisher = {American Physical Society},
    author = {Dai, Pengcheng and Hwang, H. Y. and Zhang, Jiandi and Fernandez-Baca, J. A. and Cheong, S.-W. and Kloc, C. and Tomioka, Y. and Tokura, Y.},
    month = apr,
    year = {2000},
    pages = {9553--9557},
}

@article{manDirectObservationMagnonphonon2017,
    title = {Direct observation of magnon-phonon coupling in yttrium iron garnet},
    volume = {96},
    doi = {10.1103/PhysRevB.96.100406},
    number = {10},
    urldate = {2026-09-14},
    journal = {Physical Review B},
    publisher = {American Physical Society},
    author = {Man, Haoran and Shi, Zhong and Xu, Guangyong and Xu, Yadong and Chen, Xi and Sullivan, Sean and Zhou, Jianshi and Xia, Ke and Shi, Jing and Dai, Pengcheng},
    month = sep,
    year = {2017},
    pages = {100406},
}

@article{andaEffectPhononmagnonInteraction1976,
    title = {Effect of phonon-magnon interaction on the {Green} functions of crystals and their light-scattering spectra},
    volume = {9},
    doi = {10.1088/0022-3719/9/6/024},
    number = {6},
    urldate = {2026-09-15},
    journal = {Journal of Physics C: Solid State Physics},
    author = {Anda, E.},
    month = mar,
    year = {1976},
    pages = {1075},
}

@article{streibMagnonphononInteractionsMagnetic2019,
    title = {Magnon-phonon interactions in magnetic insulators},
    volume = {99},
    doi = {10.1103/PhysRevB.99.184442},
    number = {18},
    journal = {Physical Review B},
    publisher = {American Physical Society},
    author = {Streib, Simon and Vidal-Silva, Nicolas and Shen, Ka and Bauer, Gerrit E. W.},
    month = may,
    year = {2019},
    pages = {184442},
}

@article{woodsMagnonphononEffectsFerromagnetic2001,
    title = {Magnon-phonon effects in ferromagnetic manganites},
    volume = {65},
    doi = {10.1103/PhysRevB.65.014409},
    number = {1},
    journal = {Physical Review B},
    publisher = {American Physical Society},
    author = {Woods, L. M.},
    month = nov,
    year = {2001},
    pages = {014409},
}

@article{miyawakiAchievingUltrahighResolution2026a,
    title = {Achieving ultrahigh resolution with high efficiency: {Optical} design of the two-dimensional {Resonant} {Inelastic} {X}-ray {Scattering} ({2D}-{RIXS}) spectrometer at {NanoTerasu} beamline {02U}},
    volume = {97},
    doi = {10.1063/5.0322240},
    number = {6},
    journal = {Review of Scientific Instruments},
    author = {Miyawaki, Jun},
    month = jun,
    year = {2026},
    pages = {063103},
}
\end{document}